\documentstyle[prd,aps]{revtex}
\begin{document}
 \input epsf
\draft
\renewcommand{\topfraction}{0.8}
\twocolumn[\hsize\textwidth\columnwidth\hsize\csname
@twocolumnfalse\endcsname
\preprint{UH-IfA-98-44, hep-ph/xxxxxx, June, 1998}
\title {\bf   Preheating of Fermions}
\author{Patrick B. Greene}
\address{ Department of Physics and Astronomy, University of Hawaii,
2505 Correa Rd., Honolulu HI 96822, USA} 
\author{Lev Kofman}
\address{ Institute for Astronomy, University of Hawaii,
2680 Woodlawn Dr., Honolulu, HI 96822, USA }
\date { \today	}
\maketitle
\begin{abstract} 
In  inflationary cosmology, the  particles constituting the Universe are
created after inflation in the process of reheating  due to their interaction
with the oscillating inflaton field. In the bosonic sector, the leading channel
of particle production is the non-perturbative regime of parametric
resonance,  preheating, during which bosons are created exponentially fast.
Pauli blocking prohibits the unbounded creation of fermions. For this reason,
it has been silently assumed that the creation of fermions can be treated  with
perturbation theory for the decay of individual inflatons. We consider the 
production of fermions interacting with the coherently  oscillating inflatons.
We find that the actual particle  production  occurs in a regime of the
{\it parametric excitation of fermions}, leading to  preheating of fermions.
Fermion preheating differs significantly from the perturbative expectation.
It turns out that the number density of fermions varies periodically 
with time. The total number of fermions  quickly   saturates to an average
value within a broad range of momenta $\propto q^{1/4}$,  where $q$ is the 
usual resonance parameter. The  resonant excitation of fermions  may
affect the transfer inflaton energy, estimations of the 
reheating temperature, and  the abundance  of superheavy  fermions  and  
gravitinos. Back in the bosonic sector, outside of the parametric
resonance bands there is an additional effect of parametric excitation of 
bosons with bounded occupation numberin the momentum range $\propto q^{1/4}$.
\end{abstract}
 \vskip2pc]

In the inflationary scenario,
 the Universe initially expands quasi-exponentially
in a vacuum-like state with   vanishing temperature,
entropy, and particle number densities.
Consider a simple chaotic inflation
during which all energy is contained
 in the  inflaton field $\phi$ which is 
slowly rolling down to the minimum of its effective potential
$V(\phi)$.
 When inflation ends at  $\phi \sim M_p$,
the inflaton   field begins to oscillate 
near the minimum of
 $V(\phi)$ 
with a very large  amplitude,  $\phi \simeq {1 \over 10} M_p$.
This scalar field can be considered as
a coherent superposition of $\phi$-quasi-particles with zero momenta,
i.e. inflatons at  rest.

In this scenario,  all the particles constituting the
Universe are created due to their interactions 
with the  inflatons.
For a toy  model describing the interaction between inflatons and other
Bose  particles  $\chi$,
we may consider the term $ g^2 \phi^2 \chi^2$ or $g^2 \sigma  \phi \chi^2$
in the Lagrangian. For Fermi particles $\psi$ we can include 
a Yukawa coupling $ h \bar \psi \phi \psi$ to the inflaton.
If we  assume for simplicity that the bare masses
of the fields $\chi$ and $\psi$ are very small,
the inflatons will decay into ultra-relativistic particles.
If the creation of these particles is sufficiently slow,
the  particles will simultaneously
interact with each other and come to a state of thermal equilibrium
at  the reheating temperature $T_r$.
The theory of  this gradual reheating, 
developed long  ago  with the first models of inflation
\cite{Class},
was based on the perturbative theory
with respect to the coupling constants $g^2$ and $h$.
Consider a quantum  Bose field $\chi$
with the eigenfunctions $\chi_{k}(t)\, e^{ +i{{\bf k}} \cdot {{\bf x}}}$
for comoving momenta ${\bf k}$.
 The temporal
part of the eigenfunction   obeys an oscillator-like equation
with a time-dependent  frequency: $\Omega_k^2(t)= {{ k^2}\over a^2}
+ g^2\phi(t)^2$ or $g^2 \sigma \phi(t)$.
The iterative solution to the equation of motion for $\chi_{k}(t)$ 
gives the perturbation theory result
for
 the particles' occupation number:
$n_k \propto g^4  \ll 1$.
The resulting rate for the three-legs
process $\phi \to \chi\chi$ is given by
$  \Gamma_{ \phi \to \chi \chi} =  { g^4 \sigma^2\over 8
\pi m_{\phi}}$.  By the same methods, one finds for the process
$\phi \to  \psi\psi$ a rate
$\Gamma_{ \phi \to \psi \psi }  =  { h^2 m_{\phi}\over 8 \pi}$.
Each of the created quanta carries away 
half of an inflaton's energy, $m_{\phi} \over 2$. 
This perturbative result can be interpreted as   
individual inflatons decaying independently of each 
other into  pairs of $\chi$ or $\psi$-particles.  

However, it was recognized that
reheating can begin with  a stage of  boson production
in a regime of parametric resonance
\cite{KLS}. Indeed,  in the equation for the
eigenmode $\chi_{k}(t)$, inflatons
 act  not as individual particles  but as a coherently oscillating
field $\phi(t)$. The smallness of the coupling $g^2$ alone
does not necessarily correspond to small occupation number $n_k$. 
The oscillating effective frequency  $\Omega_k(t)$
 results in  the parametric amplification
of the modes: $\chi_k \sim e^{\mu_k t}$. In this case, the energy of 
the inflatons  is  explosively 
transferred to the bose field. This process of bosonic 
{\it preheating} occurs  very rapidly and 
far from thermal equilibrium.

For fermions, the Pauli exclusion principle prohibits the 
occupation number from exceeding $1/2$.  For this reason,
it has  been assumed that  fermions are created 
in the perturbative regime with the production rate  
$\Gamma_{ \phi \to \psi \psi }$
and occupation  number $ n_k  \propto h^2 \ll 1$,
as described in the classical papers \cite{Class}.

Let us, however, consider the Dirac equation for a
massless quantum Fermi field $\psi(t, \vec x)$:
\begin{equation}
\left[ i \gamma^{\mu} \nabla_{\mu} -h \phi(t)\right] \psi=0 \ .
\label{5}
\end{equation}
Here, the  inflatons producing fermions  also 
 act not as individual particles  but as a coherently oscillating
field $\phi(t)$. Therefore, we should check 
whether coherently acting inflatons produce
the same number of fermions as an ensemble of individually decaying inflatons.
It turns out that the actual production of fermions from the oscillating
inflaton field dramatically differs from the 
perturbative approximation.

We begin our investigation with the model 
 ${1 \over 4}\lambda \phi^4 + h\bar \psi \phi \psi$.
A special feature of this theory is that
the problem of fermion production by the inflaton $\phi$
in an expanding universe can be 
 completely reduced to a similar 
problem in Minkowski space-time.
Indeed, let us perform a conformal transformation of the involved
 fields, $\varphi \equiv a \phi$ and $\Psi \equiv a^{3/2}\psi$, 
and use a conformal  time variable,
 $\tau \equiv  \sqrt{\lambda   \tilde\varphi^2 }\int {dt\over a(t)}$.
Then (without decay) the amplitude of the transformed field $\varphi$
is constant,  $\tilde\varphi$.
The background inflaton solution  is given by an elliptic function
$\varphi(\tau) \approx   \tilde\varphi  ~
  cn \left( \tau, { 1 \over \sqrt{2}}\right)$ and
the period of the background oscillations $T \approx 7.416$ 
in this model \cite{GKLS}.

The equation  for the eigenfunctions of the 
quantum fluctuations in this theory 
 can be reduced to a second-order equation for an
auxillary field $X(\tau, \vec x)$,
so that
 $\Psi= \left[ i \gamma^{\mu} \nabla_{\mu} +h \varphi\right]
 X$. The eigenmodes of the auxillary field have the form
$ X_k(\tau)e^{ +i{{\bf k}} \cdot {{\bf x}}}R_r$, 
where the $R_r$ are  eigenvectors
of the Dirac matrix $\gamma^0$ with eigenvalue $+1$.
The temporal part of the eigenmode
obeys an oscillator-like equation with a complex frequency
which depends periodically on time
\begin{equation}
\ddot X_k  +  {\left(\kappa^2  +
 q f^2 -i   \sqrt{q} \dot f 
 \right)} X_k  = 0 \ .
\label{10}
\end{equation}
The comoving momentum $k$ enters the equation in the combination
 ${ k^2 \over  \lambda   \tilde\varphi^2} \equiv \kappa^2$;
therefore, the natural units of momentum are
$\sqrt{\lambda }  \tilde\varphi$.
The background oscillations enter in the form
 $f(\tau)=cn \left( \tau, { 1 \over \sqrt{2}}\right)$ having
unit amplitude.
The combination of the coupling parameters ${h^2 \over \lambda} \equiv q$
ultimately defines the solutions to Eq.~(\ref{10}).

Equation (\ref{10}) will be the master equation in our investigation.
We choose the vacuum  positive-frequency initial condition
$X_k(\tau_0) ={\cal N}_ke^{-i  \kappa \tau}$.
The normalization factor is
$ {\cal N}_k=\left(2\Omega_k(\Omega_k+\sqrt{q} f)\right)^{-1/2}$
with the real part of the effective frequency given by
$\Omega_k^2 \equiv \kappa^2 +q f(\tau)^2$.
Notice that the problem of the production of bose particles $\chi$
in the theory ${1 \over 4}\lambda \phi^4 + {1 \over 2}g^2 \phi ^2 \chi^2$
is reduced to an equation similar to (\ref{10}) but without the
imaginary part of the frequency and with the resonance
parameter  $q={g^2\over \lambda}$  \cite{GKLS}.

Using the  standard  $\Psi$-field operator expansion, we can express
the comoving occupation number of  particles in a given spin state
through the solution of Eq.~(\ref{10}) (see, e.g., \cite{dk} for details)
\begin{equation}
n_k(\tau)={1\over 2}- 
{\kappa^2 \over \Omega_k} {\it Im} \left(X_k \dot X_k^{*} \right)
-{\sqrt{q} f \over {2\Omega_k}} \ .
\label{11}
\end{equation}
Certainly $n_k \leq 1/2$ .
The total comoving number density of particles and antiparticles
in both spin states is  ${ 4 \over (2\pi)^3 }\int d^3k~ n_k$.
Hence, the energy density of created fermions is
 $ \epsilon_{\psi}={ 1\over 2\pi^3 }\int d^3k~ \Omega_k~ n_k$.

\begin{figure}[t]
\centering
 \hskip -0.5 cm
\leavevmode\epsfysize= 5.5cm \epsfbox{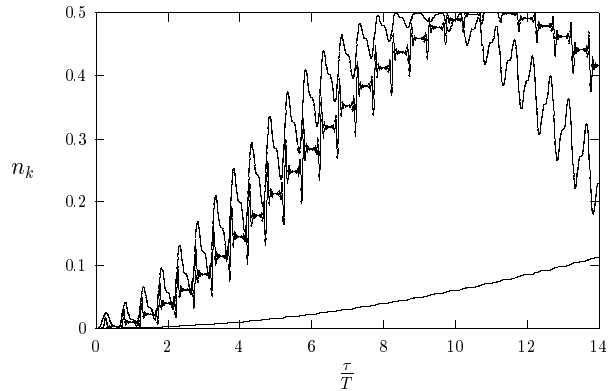}\\
\caption[fig1]{\label{fig:1}
The occupation number $n_k$ 
in $\lambda \phi^4$-inflation 
as a function of time $\tau$
(in units of $T$)  for 
$q \equiv {h^2 \over \lambda} = 10^{-4}$ (lower), $1$ (middle on right), 
and $100$ (upper on right) and
$\kappa^2 = 0.18, 1.11$, and $11.9$, respectively.
The period of the  modulation 
$\pi \over {\nu_k}$
(see Eq.~(\ref{average}))  is about $88$, $20$
and  $22$ (in units of $T$)   accordingly.}
\end{figure}

Eq.~(\ref{10}) can be solved numerically and the occupation number of
particles can be found with (\ref{11}). The  results for $ n_k(\tau)$
with different parameters $q$ are plotted in Fig.~\ref{fig:1}.

The occupation number exhibits 
high frequency (period $< {T \over 2}$)
oscillations which are modulated by a long period behavior.
If we average the occupation number over these high frequency 
oscillations, 
$\bar n_k(\tau)= { 1 \over T} \int_{\tau}^{(\tau+T)} d \tau  n_k(\tau)$,
we can write the smoothed occupation number of fermions
in a factorized form
\begin{equation}
\bar n_k(\tau) = F_k \sin^2  \nu_k \tau \ .
\label{average}
\end{equation}
It is remarkable that, for all $q$, the occupation number
of fermions evolves {\em periodically} with time \cite{time}.
The spectrum (envelope) function $F_k$
and the period ${\pi \over \nu_k}$ depend on the parameter $q$.
Typically ${\pi \over \nu_k} \sim 10 - 100$, see Fig.~\ref{fig:4}.
Therefore, the fermionic modes  get fully exited around their
time-average value $F_k \over 2$
within just tens of inflaton oscillations!
This is very different from the result of the  perturbative
approximation which gives the timing for  excitation $\sim 
\Gamma_{ \phi \to \psi \psi }^{-1}$ which is about $10^{14}$
inflaton oscillations \cite{KLS97}.
Notice that for $q \gg 1$, the number of fermions is almost constant
between two successive zeros of the inflaton field. It jumps 
in a step-like manner at instances when $\phi(t)$ crosses zero.
This is a familiar
picture for the creation of bosons explained in \cite{KLS97,GKLS}.
However, contrary to the bosonic resonance which is unbounded, 
the fermionic resonance reaches its peak value $F_k \le 1/ 2$
after a time $ \pi \over {2 \nu_k}$ and then falls back to zero,
where it begins the cycle again.

Fortunately, there is a semi-analytic shortcut for the calculation of
$F_k $ and $\nu_k$ \cite{factor}.
We found that the envelope function is given by the expression
\begin{equation}
F_k={1 \over {\sin^2 \nu_k T}} \,
{\kappa^2 \over {2 \Omega_k^2}} \,  \left({\it Im} X_k^{(1)}(T)\right)^2  \ .
\label{15}
\end{equation} 
Here, $ X_k^{(1)}(T)$
is the value of the first
fundamental solution of Eq.~(\ref{10})
(defined by the intial conditions 
$ X_k^{(1)}(0)=1$, 
$ \dot X_k^{(1)}(0)=0$) taken
after one full oscillation;
it is a function of $k$ which depends on the
parameter $q$. The modulation frequency $\nu_k$ is 
given by the relation $\cos \nu_k T = -{\it Re} X_k^{(1)}(T)$.
Therefore, to find $F_k$ and $\nu_k$, 
one need only calculate the complex value 
$ X_k^{(1)}(T)$ instead of performing a full numerical
integration of Eq.~(\ref{10}).

\begin{figure}[t]
\centering
 \hskip -0.5 cm
\leavevmode\epsfysize= 5.5cm \epsfbox{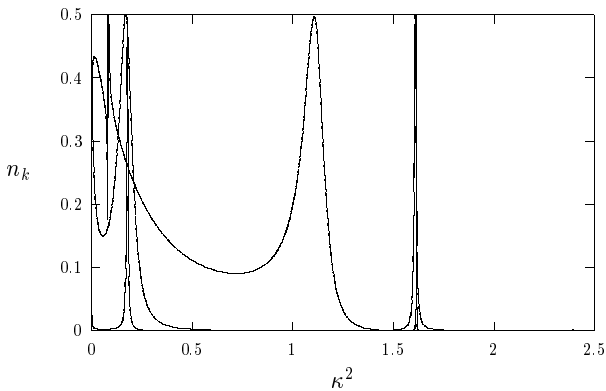}\\
\caption[fig1]{\label{fig:2}
The envelope functions $F_k$ showing the bands of fermion
resonance excitation in $\lambda \phi^4$-inflation for   
$q \equiv {h^2 \over \lambda} = 10^{-4}, 10^{-2}$, and $1.0$
(the narrowest to broadest band, respectively).   
The band in the case $q = 10^{-2}$ already deviates considerably
from the perturbative expectation. }
\end{figure}

\begin{figure}[t]
\centering
\leavevmode\epsfysize=5.3cm \epsfbox{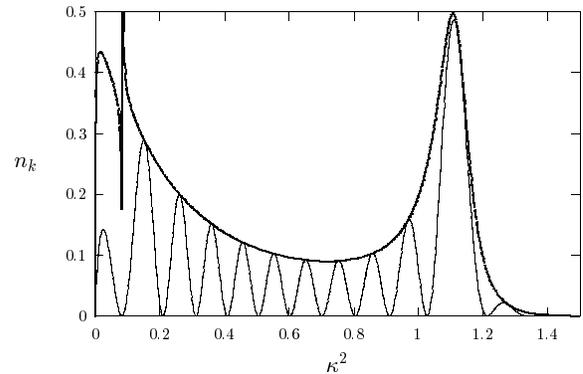}\\

\caption[fig1]{\label{fig:fig1a} 
The resonance excitation band of fermions in $\lambda \phi^4$-inflation for 
$q \equiv {h^2 \over \lambda} = 1.0$.  The heavy line shows
the envelope $F_k$ of maximum occupation number calculated from
Eq.~(\ref{15}).  The light curve is the actual occupation
number  $n_k$ of each mode, calculated from Eq.~(\ref{average}),
 after $10$  background oscillations.}
\end{figure}

\begin{figure}[t]
\centering
\hskip -0.5 cm
\leavevmode\epsfysize= 5.5cm \epsfbox{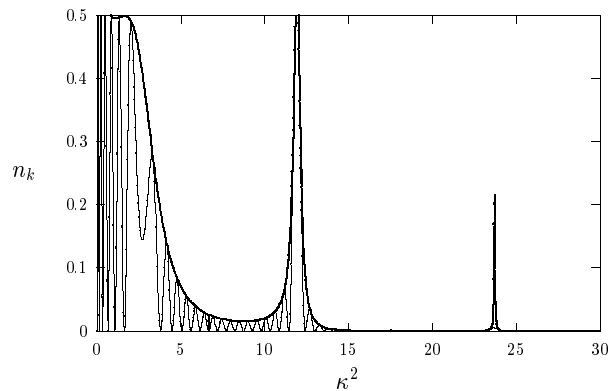}\\
\caption[fig1]{\label{fig:1c}
As in Fig.~3,
the resonance excitation band in $\lambda \phi^4$-inflation for 
$q \equiv {h^2 \over \lambda} = 100.0$.  The heavy line shows
the envelope $F_k$.
  The light curve is  the actual occupation number of
each mode $n_k$ after $10$ background oscillations.}
\end{figure}

\begin{figure}[t]
\centering
\hskip -0.5 cm
\leavevmode\epsfysize= 5.5cm \epsfbox{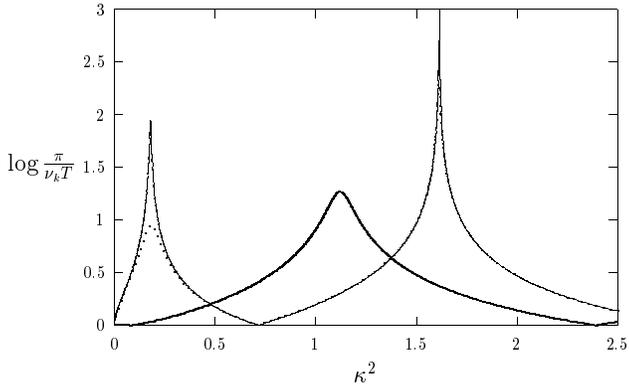}\\
\caption[fig1]{\label{fig:4}
The $\log$ of the period of modulation ${\pi \over \nu_k} $
(in units of $T$) as a function of $\kappa^2$ for 
$q \equiv {h^2 \over \lambda} = 10^{-4}, 10^{-2}$, and $1.0$
for the light, dotted, and heavy  curves respectively.}
\end{figure}

We calculated $ X_k^{(1)}(T)$ numerically and
constructed the envelope  function $F_k$ 
plotted in  Fig.~\ref{fig:2}.
The spectrum of $n_k$ is very different from what is
expected from the perturbative calculations, where
it is narrowly peaked around $\kappa={ \pi \over T}\approx 0.424 $
with the width $\Gamma_{ \phi \to \psi \psi }^{-1}$.
In  Figs.~\ref{fig:fig1a}-\ref{fig:1c} we show,
using  (\ref{15}), how the 
fermionic resonance bands are filled after
$10$ background oscillations. 

The function $\nu_k$ gives us the time scale
for fermion excitation.
In Fig.~\ref{fig:4} we plot the period
of modulation $\pi \over \nu_k$ as a function of $k$.
This function is peaked where $F_k$ is peaked, i.e.
the peaks of the resonance curve are the last to fill.
For $q \gtrsim 1$, the period for modes in the 
principle  peak is about $20 T$.
When $q \ll 1$, modes in the principle peak grow more slowly, 
with a period of approximately  $ { T \over \sqrt{q}}$.
Interestingly,  the period of modes out of the
$F_k$-peaks is significantly shorter.
For this reason, the integrated number of fermions is quickly 
saturated to its time-average value.


The huge differences in the spectrum and rate of
excitation of fermions produced by the coherently
oscillating background field vs. those in the
naive perturbative calculations are due to the
interference of the fermion fluctuations.
In this respect we can talk about 
{\it the parametric excitation of fermions}, where
$q$ plays the role of the resonance parameter,
and the  frequency $\nu_k$ can be viewed as the fermionic conterpart of
the bosonic characteristic exponent $\mu_k$
 \footnote{Notice that, for the resonant bands in the bosonic case,
the  modulating factor in Eq.~(\ref{average})
is replaced by ${\sinh^2\mu_k t }$.
This leads to the  familiar estimation
$n_k \sim e^{2\mu_k t}$ for the number of bosons.}.
The amplitude of $n_k$ shows a distinct sequence of resonance 
bands whose widths sharply narrow with increasing zone number. 
This is again reminiscent of the usual bosonic resonance. 
For $q \ll 1$, the resonance bands are located at 
$\kappa = {\pi \over T} (2l+1)$, where $l=0, 1, 2,$ \ldots.

An important issue is the width of the parametric
excitation of fermions.  From Fig.~\ref{fig:2}
we notice that for any $q$ the first, broadest zone has the 
width $\Delta \kappa \sim q^{1/4}$.
This can be understood from the
condition for non-adiabatic excitation
of particles, $\dot \Omega_k \geq \Omega_k^2$,
which gives us $\Delta   k  \leq q^{1/4}
\sqrt{\lambda} \tilde \varphi $.
This is similar to how the result was elaborated for 
the broad bosonic resonance \cite{KLS97}.
There are also very narrow zones with higher momenta,
which are also  quickly saturated.
This may be important for the production of
superheavy fermions with a mass $m_{\psi} \sim 
q^{1/4} m_{\phi}$, where   the
effective inflaton mass  $m_{\phi} \simeq 10^{13}$ Gev.

Let us estimate the energy that will be accumulated in
created fermions. Averaging over short-period oscillations, 
we have 
\begin{equation}
\bar \epsilon_{\psi}={ 2 \over \pi^2} \int dk\, k^2 \, F_k \, \sin^2
 \nu_k \tau \ .
\label{energy}
\end{equation}
For the most interesting case of the broad  resonance excitation, $q \gg 1$,
we get $\epsilon_{\psi} \sim 0.1 h^2 q^{1/4} \epsilon_{\phi}$, where
 the inflaton energy is
 $ \epsilon_{\phi}={1 \over 4} \lambda \tilde \varphi^4$.
In chaotic ${1 \over 4} \lambda \phi^4$-inflation,
 $\lambda \simeq 10^{-13}$.
If the resonance parameter  $q$ is large but the coupling parameter
is small, $h \leq 0.1$,
then only a small fraction of the inflaton energy will be 
 transferred into fermions.
However, if we push the parameters to $h \gtrsim 0.3$ ($q \gtrsim 10^{12}$),
the inflaton energy can be transferred  into fermions.
This is quite an extreme case, because it requires SUSY  cancellation
of the radiative corrections $\sim h^4 \phi^4 $. 
However, in the hybrid inflation scenario,  for example, 
there is more freedom to chose the model parameters which control the 
efficiency
of preheating \cite{hybrid}. There may be a
 range of parameters for which the inflaton energy is fully converted into
fermions even before the bosonic resonance develops.

\begin{figure}[t]
\centering
 \hskip -0.5 cm
\leavevmode\epsfysize= 5.5cm \epsfbox{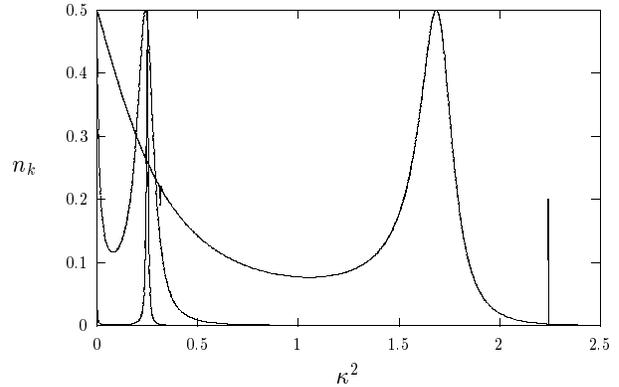}\\
\caption[fig1]{\label{fig:3}
The resonance band in $m^2 \phi^2$-inflation with  
$q \equiv {h^2 {\Phi^2} \over {m^2}} = 10^{-4}, 10^{-2}$, and $1.0$
for the narrowest to broadest band, respectively.   
The band in the case $q = 10^{-2}$ already deviates considerably
from the perturbative expectation.}
\end{figure}

Let us now turn to the model ${1 \over 2} m ^2\phi^2 + h \bar \psi \phi \psi$,
where $m$ is the inflaton mass. 
In this inflationary model the inflaton evolution  is simply
described by harmonic oscillations, $\phi \approx \Phi(t) \sin mt$,
with decreasing amplitude,
$\Phi(t)   ={ M_p \over {\sqrt{3\pi}mt}} $.
However, the problem of fermion production in this model cannot be reduced
to a problem in Minkowski space-time; the expansion of the
universe is essential. 
Let us assume, for a moment, that there is no expansion and the amplitude
$\Phi(t) =const$. 
Then the master equation would 
be Eq.~(\ref{10}) where now the resonance parameter 
$q \equiv {h^2 {\Phi^2} \over {m^2}}$ and 
$\kappa^2 \equiv {{k^2} \over {m^2}}$ (c.f. the bosonic resonance
 \cite{KLS97}). 
The result for the occupation number of fermions in this case
can also be found with  Eq.~(\ref{15}) and is plotted 
in Fig.~\ref{fig:3}.

Without expansion of the universe,  the qualitative result is similar
to that for the conformal theory. However, if we take the expansion
 into account,
we can expect an adiabatic change in the structure of the fermion
parametric excitation. This is similar to the erasure of differences between
resonant and non-resonant modes  and the onset of stochastic resonance
in the bosonic case \cite{KLS97}. Therefore, we conjecture that,
for the massive inflaton, the parametric excitation of fermions
fills a Fermi-sphere of radius 
 $\Delta k \sim  q^{1/4} m$ without  distinct resonance zones.

If expansion redshifts the fermion modes  fast enough,  it 
can prevent them from being resonantly excited.   We found that
 for $q_0 \leq 1$ or $h \leq 10^{-6}$   we return to the
perturbation theory result for the rate of fermion pair  production
${ 1 \over a^{4}} {d \over dt} ( a^4 n_{\bar \psi \psi})
 \approx \Gamma_{ \phi \to \psi \psi } n_{\phi}$,
where the comoving number density of inflatons is $n_{\phi}=m \Phi^2 a^3/2$.
This is to be interpreted as the creation of a pair of 
fermions,  $\bar \psi$ and  $\psi$,
from the decay of a single inflaton
at the instant of resonance between a given momentum 
$k$ and inflaton mass $m$, $k= { 1 \over 2}m a(t)$.

In this paper we studied the production  of fermions by an oscillating inflaton
 field, and found that  fermions are created in a regime of
 parametric excitation, which is very different from the perturbative regime.
This can be important to many interesting cosmological applications including
the channel of  inflaton energy transfer,
the estimation of the reheating temperature, 
the production of superheavy fermions, 
the production of gravitinos in the supergravity inflationary models
 \cite{gravitino}, etc.
It also elucidates another feature of bosonic preheating, missed in the
previous studies. Indeed, in the resonance bands bosons are exponentially
unstable, $n_k \sim e^{2\mu_k t}$. However, outside of the resonance bands
$\mu_k$ is imaginary and we will have bounded excitation of
bosonic fluctuations $n_k \sim O(1)$  given by Eq.~(\ref{average}).
How this will  alter the preheating scenario and its lattice simulation,
based on classical approximation of fluctuations \cite{numer},
shall be seen.

The authors are grateful to Andrei Linde for fruitful discussions,
and organizers of the CAPP98 Workshop at CERN, June 7-12,
where the results of the paper were reported
\footnote{See 
{\rm http://wwwth.cern.ch/capp98/programme.html} for transparencies.}.
This work was supported  by NSF grant AST95-29-225.
After this work was completed, we learned about related report  \cite{Baacke}.

\end{document}